\documentclass[pra,twocolumn,10pt,epsfig,graphics,showkeys,showpacs,floatfix,mathbbm]{revtex4}
\usepackage{amsmath,times}
\usepackage{psfrag}
\begin{document}
\title{Scheme for implementing quantum information sharing via
tripartite entangled state in cavity QED}
\author{Zheng-Yuan Xue\footnote{E-mail: xuezhengyuan@yahoo.com.cn},
You-Min Yi and Zhuo-Liang Cao\footnote{E-mail: zlcao@ahu.edu.cn
(Corresponding author).}}

\affiliation{Key Laboratory of Opto-electronic Information
Acquisition and Manipulation Ministry of Education, School of
Physics {\&} Material Science, Anhui University, Hefei, 230039,
People's Republic of China}
\begin{abstract}
We investigate economic protocol to securely distribute and
reconstruct a single-qubit quantum state between two users via a
tripartite entangled state in cavity QED. Our scheme is insensitive
to both the cavity decay and the thermal field.
\end{abstract}


\maketitle

\section{introduction}
Entanglement is one of the most counterintuitive features in quantum
mechanics; assisted with entangled state one can complete many
impossible tasks within the classical world. One of the most
striking applications of entanglement is quantum secret sharing
(QSS) \cite{Hillery}. It is a process to distribute private key
among three \cite{Hillery} or multiply parties \cite{Hillery,xia}
securely. If and only if when they cooperate, they can get complete
information about the message. Meanwhile, if one of them is
dishonest, the honest players may keep the dishonest one from doing
any damage. But, up to now, most existing QSS schemes only focused
on creating a private key or splitting a classical secret among many
parties. Recently, due to its promising applications in quantum
secure communication, it attracts many attentions. However, many
applications in quantum information theory require the distribution
of quantum states. Cleve et al. \cite{Cleve} proposed a protocol
providing robust and secure distribution of quantum states between
nodes, which Lance et al. \cite{Lance 1} termed as \textit{quantum
state sharing} to differentiate from the QSS of classical
information. As quantum state carries quantum information, it is
also called \textit{quantum information sharing} (QIS). Quantum
state sharing is a protocol where perfect reconstruction of quantum
state is achieved with partial information of a multipartite quantum
network. It also allows for secure communication in a quantum
network where partial information is lost or acquired by disloyal
parties.

Recently, Lance et al. demonstrated schemes for encoding a secret
coherent state into a tripartite entangled state and distributing it
to two players both theoretically \cite{Lance 1} and experimentally
\cite{Lance 2}. Deng et al. \cite{Deng}, proposed a scheme for
multiparty quantum state sharing of an arbitrary two-particle
entanglement with EPR pairs. Li et al. \cite{Li} investigated
protocol for multiparty secret sharing of single-qubit quantum
information.  In the previous schemes, Bell state measurements (in
Ref. \cite{Deng}) or even multipartite joint measurements (in Ref.
\cite{Li}), which is still under extensive research among many
ambitious scientists, are widely employed. Zhang et al.
\cite{Zhang}, proposed a scheme for quantum information sharing of a
single-qubit state based on entanglement swapping in cavity QED
where the effects of cavity decay and thermal field are both
eliminated. However, in quantum information theory, entangled states
are precious resource, thus one should employ them with great
deliberation. In Ref. \cite{Zhang}, they utilize an EPR and
tripartite GHZ state (five atomic qubits) to finish the task of a
single-qubit state sharing between two parties while four qubits
(two EPR states) can complete the task \cite{Deng}. Furthermore Ref.
\cite{Hillery} showed that three qubits (a pure tripartite entangled
GHZ type state) is sufficient.

It is well known that multipartite qubits can be entangled in
different inequivalent ways. For tripartite entangled quantum
system, it falls into two classes of irreducible entanglements
\cite{entangle}, that is, GHZ and W class state. The motivation of
classifying entangled state is that, if the entanglement of two
states is equivalent, then the two states can be used to perform the
same task, although the probability of successful performance of the
task may depend on the amount of entanglement of the state. But, in
the branch of quantum state sharing, as well as in QSS, most of the
previous schemes \cite{Hillery,xia,Cleve,Lance 1,Lance
2,Li,Deng,Zhang} utilize the GHZ class of entangled state. W class
state is also a promising candidate in implementing quantum
communication and other tasks in the realm of quantum information
processing. Recently, Joo et al. \cite{joo} presented a novel scheme
for  secure quantum communication via W state, where they proposed
three different protocols for secure quantum communication, that is,
quantum key distribution, probabilistic quantum secret sharing of
classical information and their synthesis. In order to extensively
investigate the applications of W class states in quantum
communication, we engage ourselves in the work of implementing
quantum state sharing via W class state in this paper. Here, we
first investigate a physical scheme for Ref. \cite{Hillery} in
cavity QED, which shares single-qubit quantum state via a tripartite
GHZ state. Then we will argue that W class state can also fulfill
the task probabilistically. The distinct advantage of the scheme is
that during the passage of the atoms through the cavity field, a
strong classical field is accompanied, thus our scheme is
insensitive to both the cavity decay and the thermal field.

\section{The cavity model}
We consider two identical two-level atoms simultaneously interacting
with a single-mode cavity and driven by a classical field. Then the
interaction between the single-mode cavity and the atoms can be
described, in the rotating-wave approximation, as \cite{ham}
\begin{eqnarray}
 H=\omega_{0} S_{Z} +\omega_{a} a^{+}a +
 \sum_{j=1}^{2}[g(a^{+}S_{j}^{-}+aS_{j}^{+})\nonumber\\
+\Omega(S_{j}^{+}e^{-i\omega t}+S_{j}^{-}e^{i\omega t})],
\end{eqnarray}
where $S_{Z}=\frac{1}{2}\sum_{j=1}^{2} (|e\rangle_{j,j}\langle
e|-|g\rangle_{j,j}\langle g|)$, $S_{j}^{+}=|g\rangle_{j,j}\langle
e|$, $S_{j}^{-}=|e\rangle_{j,j}\langle g|$ and $|e\rangle_{j}$,
$|g\rangle_{j}$ are the excited and ground states of  $j$th atom,
respectively. $a^{+}$ and $a$ are the creation and annihilation
operators for the cavity mode, respectively. $g$ is the coupling
constant between atomic system and the cavity, $\Omega$ is the Rabi
frequency, $\omega_{0}$ , $\omega_{a}$ and $\omega$ are atomic
transition frequency, cavity frequency and the frequency of the
driven classical field, respectively.

While in the case of $\omega_{0}=\omega$, in the interaction
picture, the interaction Hamiltonian is
\begin{eqnarray}
\label{abc1} H_{I}=\Omega\sum_{j=1}^{2}(S_{j}^{+}+S_{j}^{-})
+g\sum_{j=1}^{2} (e^{-i\delta t}a^{+}S_{j}^{-}+e^{i\delta
t}aS_{j}^{+}),
\end{eqnarray}
where $\delta$ is the detuning between atomic transition frequency
$\omega_{0}$  and the cavity frequency $\omega_{a}$.

We define the new atomic basis
\begin{eqnarray}
|+\rangle_j=\frac{1}{\sqrt{2}}(|g\rangle_j+|e\rangle_j),
|-\rangle_j=\frac{1}{\sqrt{2}}(|g\rangle_j-|e\rangle_j).
\end{eqnarray}
Then we can rewrite $H_{I}$ as
\begin{eqnarray}
H_{I}=\sum_{j=1}^{2}[2\Omega\sigma_{z,j}+ge^{-i\delta
t}a^{+}(\sigma_{z,j}+\frac{1}{2}\sigma_{j}^{+}-\frac{1}{2}\sigma_{j}^{-})\nonumber\\
+ge^{i\delta
t}a(\sigma_{z,j}+\frac{1}{2}\sigma_{j}^{-}-\frac{1}{2}\sigma_{j}^{+})],
\end{eqnarray}
where
$\sigma_{z,j}=\frac{1}{2}(|+\rangle_{j,j}\langle+|-|-\rangle_{j,j}\langle-|)$,
$\sigma_{j}^{+}=|+\rangle_{j,j}\langle-|$ and
$\sigma_{j}^{-}=|-\rangle_{j,j}\langle+|$.

The time evolution of this system is decided by Schr\"{o}dinger
equation
\begin{eqnarray}
i[d|\psi(t)\rangle/dt]=H_I|\psi(t)\rangle.
\end{eqnarray}
We perform the unitary transformation
$|\psi(t)\rangle=exp(-iH_{0}t)|\psi(t)^{'}\rangle$ with
$H_0=2\Omega\sum_{j=1}^{2}\sigma_{z,j}=\sum_{j=1}^{2}\Omega
(S_{j}^{+}+S_{j}^{-})$, then we obtain
\begin{eqnarray}
i[d|\psi(t)^{'}\rangle/dt]=H_{I}^{'}|\psi(t)^{'}\rangle,
\end{eqnarray}
with
\begin{eqnarray}
H_{I}^{'}=\sum_{j=1}^{2}[ge^{-i\delta
t}a^{+}(\sigma_{z,j}+\frac{1}{2}\sigma_{j}^{+}e^{2i\Omega
t}-\frac{1}{2}\sigma_{j}^{-}e^{-2i\Omega t})\nonumber\\
+ge^{i\delta
t}a(\sigma_{z,j}+\frac{1}{2}\sigma_{j}^{-}e^{-2i\Omega
t}-\frac{1}{2}\sigma_{j}^{+}e^{2i\Omega t})].
\end{eqnarray}

In the strong driving regime ($2\Omega \gg\delta, g$), we can
realize a rotating-wave approximation, \emph{i.e}., eliminate the
terms oscillating fast, which induce Stark shifts. By doing so one
just introduce a ignorable imperfection of the generated state
\cite{effham}. Then the interaction Hamiltonian reduces to
\begin{eqnarray}
H_{I'}&&=g\sum_{j=1}^{2} (e^{-i\delta t}a^{+}+e^{i\delta
t}a)\sigma_{z,j}\nonumber\\
&&=\frac{g}{2}\sum_{j=1}^{2} (e^{-i\delta
t}a^{+}+e^{i\delta t}a)(S_{j}^{+}+S_{j}^{-}).
\end{eqnarray}

In the case of large detuning (2$\delta \gg g$), there is no energy
exchange between the atomic system and the cavity. The resonant
transitions are
$|e\rangle_{j}|g\rangle_{k}|n\rangle\leftrightarrow|g\rangle_{j}|e\rangle_{k}|n\rangle$
and
$|e\rangle_{j}|e\rangle_{k}|n\rangle\leftrightarrow|g\rangle_{j}|g\rangle_{k}|n\rangle$.
The transition
$|e\rangle_{j}|g\rangle_{k}|n\rangle\leftrightarrow|g\rangle_{j}|e\rangle_{k}|n\rangle$
is mediated by $|e\rangle_{j}|e\rangle_{k}|n\pm1\rangle$ and
$|g\rangle_{j}|g\rangle_{k}|n\pm1\rangle$. The contributions of
$|e\rangle_{j}|e\rangle_{k}|n\pm1\rangle$ are equal to those of
$|g\rangle_{j}|g\rangle_{k}|n\pm1\rangle$, and the corresponding
Rabi frequency is $g^{2}/2\delta$. Since the transition paths
interfere destructively, the Rabi frequency is independent of the
photon number of the cavity mode. The Rabi frequency for
$|e\rangle_{j}|e\rangle_{k}|n\rangle\leftrightarrow|g\rangle_{j}|g\rangle_{k}|n\rangle$,
mediated by $|e\rangle_{j}|g\rangle_{k}|n\pm1\rangle$ and
$|g\rangle_{j}|e\rangle_{k}|n\pm1\rangle$, is also $g^{2}/2\delta$.
The Stark shift for the states $|e\rangle_{j}$ and $|g\rangle_{j}$
are both equal to $g^{2}/4\delta$. The photon-number-dependent Stark
shifts $g(e^{-i\delta t}a^{+}S_{j}^{+}+e^{i\delta t}aS_{j}^{-})$
induced by the strong classical field are negative to those induced
by $g(e^{-i\delta t}a^{+}S_{j}^{-}+e^{i\delta t}aS_{j}^{+})$, thus
the photon-number-dependent Stark shifts are also canceled. Then the
effective interaction Hamiltonian can be described as\cite{effham}
\begin{eqnarray}
H_{e}&&=\frac{\lambda}{2}[\sum_{j=1}^{2}(|e\rangle_{j,j}\langle
e|+|g\rangle_{j,j}\langle g|)\nonumber\\
&&+\sum_{j,k=1;j\neq
k}^{2}(S_{j}^{+}S_{k}^{+}+S_{j}^{+}S_{k}^{-}+H.c.)],
\end{eqnarray}
where $\lambda=g^{2}/2\delta$. The distinct feature of the effective
Hamiltonian is that it is independent of the photon number of the
cavity field, that is, allowing it to be in a thermal state. Our
scheme is based on such kind of cavity, so it is insensitive to both
the cavity decay and the thermal field. The time evolution operator
of the system is
\begin{eqnarray}
\label{u} U(t)=e^{-iH_{0}t} e^{-iH_{e}t}.
\end{eqnarray}

\section{QIS via GHZ state}
Suppose Alice wants to send quantum information (a single-qubit
state) to a remote user, it is
\begin{equation}
\label{abc} |\psi\rangle_{1}=(\alpha|e\rangle+\beta|g\rangle)_{1},
\end{equation}
where $\alpha$,  $\beta$ are unknown coefficients and
$|\alpha|^{2}+|\beta|^{2}=1$. As she doesn't know whether he /she is
honest, she makes the quantum information shared between two users.
If and only if they collaborate, both users can reconstruct the
state on their atom, in addition, individual users could not do any
damage to the process. Here, we assume that communication over a
classical channel is insecure, which means we can't resort to the
simplest method of teleportation \cite{Bennett 1}. One could also
securely complete the task using standard quantum cryptography
\cite{Bennett 2}, which requires more resource and measurements
\cite{Hillery}. Initially, Alice possesses three atoms and they are
prepared in the GHZ state
\begin{equation}
|\psi\rangle_{2,3,4}=\frac{1}{\sqrt{2}}(|eee\rangle+i|ggg\rangle)_{2,3,4}
\end{equation}
The combined state of the four atoms is
\begin{equation}
|\psi\rangle=\frac{1}{\sqrt{2}}(\alpha|e\rangle_{1}+\beta|g\rangle_{1})\otimes(|eee\rangle+i|ggg\rangle)_{2,3,4}
\end{equation}
Alice simultaneously sends atoms 1 and 2 into a single-mode cavity
meanwhile they are driven by a classical field, the time evolution
of the system is governed by Eq. (\ref{u}). Choosing to adjust the
interaction time $\lambda t=\pi/4$ and modulate the driving field
$\Omega t=\pi$, lead the quantum state of the four atoms system,
after interaction, to
\begin{eqnarray}
|\psi\rangle=\frac{1}{2}[|ee\rangle_{1,2}(\alpha|ee\rangle+\beta|gg\rangle)_{3,4}\nonumber\\
-i|gg\rangle_{1,2}(\alpha|ee\rangle-\beta|gg\rangle)_{3,4}\nonumber\\
+|ge\rangle_{1,2}(\alpha|gg\rangle+\beta|ee\rangle)_{3,4}\nonumber\\
-i|eg\rangle_{1,2}(\alpha|gg\rangle-\beta|ee\rangle)_{3,4}]
\end{eqnarray}
Alice then sends atoms 3 and 4 to Bob and Charlie, respectively.
After Alice confirms that Bob and Charlie both receive one atom,
then she operates two measurements on atoms 1 and 2 in the basis
$\{|e\rangle,|g\rangle\}$, respectively. Then the state of atoms 3
and 4 collapse to one of the following unnormalized states
\begin{subequations}
\label{g}
\begin{equation}
 |\varphi\rangle_{3,4}=\frac{1}{2}(\alpha|ee\rangle_{3,4}+\beta|gg\rangle_{3,4}),
\end{equation}
\begin{equation}
\label{g2}
|\varphi\rangle_{3,4}=\frac{1}{2}(\alpha|ee\rangle_{3,4}-\beta|gg\rangle_{3,4}),
\end{equation}
\begin{equation}
|\varphi\rangle_{3,4}=\frac{1}{2}(\alpha|gg\rangle_{3,4}+\beta|ee\rangle_{3,4}),
\end{equation}
\begin{equation}
|\varphi\rangle_{3,4}=\frac{1}{2}(\alpha|gg\rangle_{3,4}-\beta|ee\rangle_{3,4}).
\end{equation}
\end{subequations}
Up to now, the quantum information (atom 1) is encoded into the
state of atoms 3 and 4, which is shared between Bob and Charlie,
thus the distribution of quantum information is completed. Each of
the two users can only obtain the amplitude information of atoms 1
by local operation and classical information available  (the result
of Alice's measurements). If and only if they cooperate, one of
them, and only one of them, can get the complete information of the
atoms for the sake of no-cloning theorem \cite{Hillery 2}. Assuming
the collapsed state is in the form of Eq. (\ref{g2}) after Alice's
measurements and then she declares the measurement result to Bob and
Charlie over a public channel. If Alice designates Charlie to
reconstruct the quantum state and Bob would like to cooperate with
him, then Charlie could obtain complete information of Alice's atoms
by local operation.

Next we will discuss the process in detail. Bob let his atom (atom
3) crosses a classical field tuned to the transitions
$|g\rangle\rightarrow|e\rangle$. Choose the amplitudes and phases of
the classical fields appropriately so that atom 3 undergoes the
transitions
\begin{eqnarray}
|e\rangle\rightarrow\frac{1}{\sqrt{2}}(|e\rangle_{3}+|g\rangle_{3}),
|g\rangle\rightarrow\frac{1}{\sqrt{2}}(|e\rangle_{3}-|g\rangle_{3}),
\end{eqnarray}
which leads the state in Eq. (\ref{g2}) into the unnormalized state
\begin{eqnarray}
\label{g2t}
|\varphi\rangle_{3,4}=\frac{1}{2\sqrt{2}}[|g\rangle_{3}(\alpha|e\rangle_{4}+\beta|g\rangle_{4})
+|e\rangle_{3}(\alpha|e\rangle_{4}-\beta|g\rangle_{4})].
\end{eqnarray}
Bob performs a computational basis measurement on his atom  and
informs Charlie the measurement result then the state of Charlie's
atom  is exactly the state of Alice's state or relate it up to a
corresponding unitary transformation. Charlie can choose the correct
unitary transformation based on Bob's computational measurement
result. Without Bob's cooperation, Charlie can only get the
measurement results of Alice from the public channel and can only
reconstruct the state with a successful probability of 1/2.
Similarly, Bob can also reconstruct the state on his atom if Charlie
chooses to cooperate with him. From Eq. (\ref{g2t}) we can see that
the successful probability of reconstruct the original state from
Eq. (\ref{g2}) is $1/4$. There is four equal probable states in Eq.
(\ref{g}), so the total probability of successfully reconstructing
the original state in both nodes reaches unit.

\section{QIS via W state}
If Alice initially possesses three atoms prepared in the W class
entangled state
\begin{equation}
 |\psi\rangle_{2,3,4}=(a|gge\rangle+b|geg\rangle+ic|egg\rangle)_{2,3,4},
\end{equation}
where  $|a|^{2}+|b|^{2}+|c|^{2}=1$, and we can assume $|a|>|b|>|c|$
without loss of generality. So, the combined state for the four
atoms is
\begin{eqnarray}
|\psi\rangle=(a|e\rangle+b|g\rangle)_{1}(a|gge\rangle+b|geg\rangle+ic|egg\rangle)_{2,3,4}.
\end{eqnarray}

Then, Alice simultaneously sends atoms 1 and 2 into a single-mode
cavity meanwhile they are driven by a classical field, the time
evolution of the system is governed by Eq. (\ref{u}). Choosing to
adjust the interaction time $\lambda t=\pi/4$ and modulate the
driving field $\Omega t=\pi$, lead the quantum state of the four
atoms system, after interaction, to
\begin{eqnarray}
|\psi\rangle=\frac{1}{\sqrt{2}}[|gg\rangle_{1,2}(\beta
a|ge\rangle+\beta b|eg\rangle+\alpha c|gg\rangle)_{3,4}\nonumber\\
-i|ee\rangle_{1,2}(\beta a|ge\rangle+\beta b|eg\rangle-\alpha c|gg\rangle)_{3,4}\nonumber\\
+|eg\rangle_{1,2}(\alpha a|ge\rangle+\alpha b|eg\rangle+\beta
c|gg\rangle)_{3,4}\nonumber\\
-i|eg\rangle_{1,2}(\alpha a|ge\rangle+\alpha b|eg\rangle-\beta
c|gg\rangle)_{3,4}].
\end{eqnarray}
Alice then sends atoms 3 and 4 to Bob and Charlie, respectively.
After Alice confirms that Bob and Charlie both receive one atom, she
operates two computational measurements on atoms 1 and 2,
respectively. The state of atoms 3 and 4 collapse to one of the
following unnormalized states
\begin{subequations}
\label{w34}
\begin{equation}
 |\varphi\rangle_{3,4}=\frac{1}{\sqrt{2}}(\beta a|ge\rangle+\beta b|eg\rangle+\alpha c|gg\rangle)_{3,4},
\end{equation}
\begin{equation}
|\varphi\rangle_{3,4}=\frac{1}{\sqrt{2}}(\beta a|ge\rangle+\beta
b|eg\rangle-\alpha c|gg\rangle)_{3,4},
\end{equation}
\begin{equation}
\label{w343} |\varphi\rangle_{3,4}=\frac{1}{\sqrt{2}}(\alpha
a|ge\rangle+\alpha b|eg\rangle+\beta c|gg\rangle)_{3,4},
\end{equation}
\begin{equation}
|\varphi\rangle_{3,4}=\frac{1}{\sqrt{2}}(\alpha a|ge\rangle+\alpha
b|eg\rangle-\beta c|gg\rangle)_{3,4}.
\end{equation}
\end{subequations}
Now the quantum information (atom 1) is encoded into the state of
atoms 3 and 4. Neither of the two users could obtain complete
information of Alice's atom by local operation and classical
information available. If and only if they cooperate, one of them,
and only one of them, can get the quantum information. Without loss
of generality, we assume the collapsed state is Eq. (\ref{w343})
after Alice's measurement, and then she declares the measurement
result to Bob and Charlie over a public channel. It can be rewrote
as
\begin{eqnarray}
|\varphi\rangle_{3,4}=\frac{1}{\sqrt{2}}[|g\rangle_{3}(\alpha
a|e\rangle_{4}+\beta c|g\rangle_{4}) +\alpha
b|e\rangle_{3}|g\rangle_{4}].
\end{eqnarray}

If Alice designates Charlie to reconstruct the quantum state,
meanwhile, Bob would like to cooperate (send the result of his
measurement to Charlie) then Charlie could obtain complete
information of Alice's atom by local operation. Next we will discuss
the process in detail.

Bob performs a computational basis measurement on his atom and
informs Charlie the measurement result. If Bob's measurement result
is $|g\rangle_{3}$, then the state of Charlie's atom is
\begin{equation}
|\varphi\rangle_{4}=\frac{1}{\sqrt{2}}(\alpha a|e\rangle_{4}+\beta
c|g\rangle_{4}).
\end{equation}
To reconstruct the initial state on atom 4 in cavity QED, Charlie
needs another single-mode high-Q optical cavity with the initial
state $|0\rangle$  and a photon detector. According to the
Jaynes-Cummings model, the Hamiltonian of the resonant
$(\delta=\omega_{0}-\omega=0 )$ interaction system between the atom
and the cavity is
\begin{equation}
\label{rh} H=\omega(a^{+}a+S_{z})+g(aS_{+}+a^{+}S_{-}).
\end{equation}
The time evolution of the interaction under the Hamiltonian in Eq.
(\ref{rh}) are
\begin{subequations}
\label{e}
\begin{equation}
|g\rangle|0\rangle\rightarrow|g\rangle|0\rangle,
\end{equation}
\begin{equation}
|e\rangle|0\rangle\rightarrow(\cos gt|e\rangle|0\rangle-\sin
gt|g\rangle|1\rangle).
\end{equation}
\end{subequations}
Sending atom 4 into the cavity and taking the interaction time
$t_{1}=1/g\arccos(|c|/|a|)$. According to Eq.(\ref{e}) we can get
the state of the quantum system after interaction as
\begin{eqnarray}
\label{4c}
|\Psi(t_{1})\rangle_{4,C}=\frac{|c|}{\sqrt{2}}(\alpha|e\rangle_{4}+\beta|g\rangle_{4})|0\rangle\nonumber\\
+\frac{\alpha
a\sqrt{|a|^{2}-|c|^{2}}}{\sqrt{2}|a|}|g\rangle_{4}|1\rangle
\end{eqnarray}

If Alice designates Bob to reconstruct the quantum state with the
cooperation of Charlie, the state of Eq. (\ref{w343}) should be
rewritten as
\begin{eqnarray}
|\varphi\rangle_{3,4}=\frac{1}{\sqrt{2}}[(\alpha
b|e\rangle_{3}+\beta c|g\rangle_{3})|g\rangle_{4} +\alpha
a|g\rangle_{3}|e\rangle_{4}].
\end{eqnarray}
After been informed the computational basis measurement result, Bob
sends his atom into the resonant cavity and choose the interaction
time $t_{2}=1/g\arccos(|c|/|b|)$ , we can get the state of the
quantum system after interaction as
\begin{eqnarray}
\label{3c}
|\Psi(t_{2})\rangle_{3,C}=\frac{|c|}{\sqrt{2}}(\alpha|e\rangle_{3}+\beta|g\rangle_{3})|0\rangle\nonumber\\
+\frac{\alpha
a\sqrt{|b|^{2}-|c|^{2}}}{\sqrt{2}|b|}|g\rangle_{3}|1\rangle
\end{eqnarray}

Finally, by detecting the cavity, the state of atom 3 (see Eq.
{\ref{3c}}) or atom 4 (see Eq. {\ref{4c}}) collapsed to the state of
Alice's qubit with a probability of $1/2|c|^{2}$. For the rest three
state in Eq. (\ref{w34}), the corresponding collapsed state will
relate Alice's state up to a corresponding unitary transformation.
So, the total probability of successful of our scheme is
$\textit{P}_{s}=2|c|^{2}$, which is decided by the smallest
superposition coefficients of the W state used as quantum channels.
We also note that the probability is equal to the probability of
successful teleportation with nonmaximally entangled W state
\cite{Joo}. So, the optimal probability for quantum state sharing
with W state is $\textit{P}=2/3$ , and then the nonmaximally
entangled W state will just be a maximally entangled sate
$(a=b=c=1/\sqrt{3})$.

\section{discussions}
We have presented a two-party quantum state sharing scheme via a
tripartite GHZ type entangled state, as well as a W type entangled
state. Now we will discuss the security of our scheme.

(1) If there is an eavesdropper who has been able to entangle an
ancilla with the quantum channel, and at some later time he can
measure the ancilla to gain information about the measurement
results of the legal users. However, Hillery \textit{et al}.
\cite{Hillery} showed that if this entanglement does not introduce
any errors into the procedure, then the state of the system is a
product of the entangled state and the ancilla. In other word, the
eavesdropper could gain nothing about the measurements on the
quantum channel from observing his ancilla. Conversely, if he does
gain some information about the two legal users' measurements, she
must inevitably introduce errors in the procedure.

(2) If one of the users is the eavesdropper, \textit{i.e}. Bob, who
wants to obtain Alice's information without the cooperation of the
third party and without being detected. If Alice designates Bob to
reconstruct the state and Charlie agrees to cooperate with Bob, Bob
can eavesdrop the state with unit successful probability without
being detected. Without the cooperation of Charlie, Bob can also
eavesdrop the state with a successful probability of 0.5. If Alice
designates Charlie to receive the state, he can also recover the
state with the help of Bob. If Bob lies his measurement results to
him then Bob gains nothing and Charlie can't obtain the correct
state deterministically.

(3) Bob can also get the qubit that Alice sends to Charlie, and
sends Charlie a qubit that he has prepared before hand. He wants to
reconstruct Alice's state without the help of Charlie. By doing so,
only when Alice designates him to reconstruct the state he can get
the state without being detected. If Alice assigns Charlie to
recover the state, then Bob's eavesdropping behavior can be
detected. Because Bob does not know Alice's measurement result thus
the qubit he sent to Charlie is not in the correct quantum state,
which will lead to the difference between the state recovered by
Charlie and the state Alice has sent. By checking a subset of the
state with Charlie publicly, the eavesdropping behavior of Bob can
be detected.

Next, we will give a brief analysis of the experimental feasibility
of our scheme. For the interaction between the optical cavity and
two two identical atoms, it is noted that the atomic state evolution
is independent of the cavity field state, so, during the process, it
does not require the transfer of quantum information between the
atoms and cavity. In addition, with the help of a strong classical
driving field the photon-number dependent parts in the evolution
operator are canceled. Thus the scheme is insensitive to the thermal
field and the cavity decay. So, the requirement on the quality
factor of the cavities is greatly loosened. In our scheme, the two
atoms must be simultaneously interaction with the cavity. But in
real case, we can't achieve simultaneousness in perfect precise.
Calculation on the error \cite{Zheng 2} suggests that it only
slightly affects the fidelity of the reconstruct state. Furthermore,
the time required to complete the process should be much shorter
than that of atom radiative time and cavity field decay. Osnaghi
\textit{et al}. \cite{Osnaghi} show that for the Rydberg atoms with
principal quantum numbers 50 and 51, the interaction time is much
shorter than the atomic radiative time. So our scheme is realizable
by using cavity QED techniques presently available.
\section{conclusion}
In summary, we have investigated an economic and experimentally
feasible scheme to securely distribute and reconstruct a
single-qubit quantum state between two parties via a tripartite
entangled state in cavity QED. If and only if when they cooperate
with each other, they can reconstruct the quantum state. Any
contempt to get complete information of the state without the
cooperation of the third party can't be succeed in a deterministic
way. The distinct advantage of the scheme is that during the passage
of the atoms through the cavity field, a strong classical field is
companied, thus our scheme is insensitive to both the cavity decay
and the thermal field. In addition, Our scheme only employs
single-qubit computational basis measurements, thus it may offer a
simple and easy way of demonstrating quantum state sharing
experimentally in cavity QED with atomic qubits. Our scheme can also
be generalized to multipartite case via multipartite entangled state
in a straight forwards way.

\begin{acknowledgments}
This work is supported by the Key Program of the Education
Department of Anhui Province under Grant No. 2006kj070A, the Talent
Foundation of Anhui University and the the Postgraduate Innovation
Research Plan from Anhui university entitled to Z.-Y. Xue.
\end{acknowledgments}


\begin{thebibliography}{99}

\bibitem{Hillery} M. Hillery, V. Buzek and A. Berthiaume, Phys. Rev. A
\textbf{59},1829 (1999).

\bibitem{xia} L. Xiao, G.-L. Long, F.-G. Deng and J.-W. Pan, Phys.
Rev. A \textbf{69}, 052307 (2004).

\bibitem{Cleve} R. Cleve, D. Gottesman and  H. K. Lo, Phys. Rev. Lett.
\textbf{83}, 648 (1999).

\bibitem{Lance 1} A. M. Lance, T. Symul, W. P. Bowen, B. C. Sanders and P. K.
Lam, Phys. Rev. Lett. \textbf{92}, 177903 (2004).

\bibitem{Lance 2} A. M. Lance, T. Symul, W. P. Bowen, B. C. Sanders, T. Tyc, T.
C. Ralph and P. K. Lam, Phys.Rev. A \textbf{71}, 033814 (2005).

\bibitem{Deng} F.-G. Deng, X.-H. Li, C.-Y. Li, P. Zhou and H.-Y. Zhou, Phys. Rev. A \textbf{72}, 044301 (2005).

\bibitem{Li} Y.-M. Li, K.-S. Zhang and K.-C. Peng, Phys. Lett. A
\textbf{324}, 420 (2004).

\bibitem{Zhang} Y.-Q. Zhang, X.-R. Jin and S. Zhang, Phys. Lett. A
\textbf{341}, 380 (2005).

\bibitem{entangle} M. Greenberger, M. A. Horne and A.
Zeilinger, Am. J. Phys. \textbf{58}, 1131 (1990); W. D\"{u}r, G.
Vidal and J.I. Cirac, Phys. Rev. A \textbf{62}, 062314 (2000); A.
Ac\'{\i}n, D. Bru{\ss}, M. Lewenstein and A. Sanpera, Phys. Rev.
Lett. \textbf{87}, 040401 (2001).

\bibitem{joo} J. Joo, J. Lee, J. Jang ang Y-J. Park, eprint
quantum-ph/0204003 (2002).

\bibitem{ham} E. Solano, G. S. Agarwal and H. Walther, Phys. Rev. Lett.
\textbf{90}, 027903 (2003); S.-B. Zheng, Phys. Rev. A \textbf{68},
035801 (2003).

\bibitem{effham} S.-B. Zheng, Phys. Rev. A \textbf{66}, 060303(R) (2002).


\bibitem{Bennett 1} C. H. Bennett, G. Brassard, C. Cr\'{e}peau, R. Jozsa, A. Peres and
W. K. Wootters, Phys. Rev. Lett. \textbf{70}, 1895 (1993).

\bibitem{Bennett 2} C. H. Bennett, G. Brassard. Proc. IEEE Int. Conf. on Computers, Systems and Signal Processing,
Bangalore (IEEE, New York), pp. 175-179 (1984); A. K. Ekert, Phys.
Rev. Lett. \textbf{67}, 661 (1991).

\bibitem{Hillery 2} M. Hillery, V. Buzek and A. Berthiaume, Phys. Rev. A
\textbf{54}, 1844 (1996).

\bibitem{Joo}  J. Joo, Y.-J Park S. Oh and J. Kim, New J. Phys. \textbf{5}, (2003) 136;
Z.-L. Cao and M. Yang, Physica A \textbf{337}, 132 (2004).

\bibitem{Zheng 2} S.-B. Zheng and G.-C. Guo, Phys. Rev. Lett. \textbf{85}, 2392
(2000).

\bibitem{Osnaghi} S. Osnaghi, P. Bertet, A. Auffeves, P. Maioli, M. Brune, J. M.
Raimond and S. Haroche, Phys. Rev. Lett. \textbf{87}, 037902 (2001).

\end{thebibliography}
\end{document}